\documentclass[manuscript]{aastex63}

\received{}
\revised{}
\accepted{}
\submitjournal{ApJ}

\shorttitle{Identification of CH on AIA/SDO images}
\shortauthors{Inceoglu et al.}

\begin{document}

\title{Identification of Coronal Holes on AIA/SDO images using unsupervised Machine Learning}

\correspondingauthor{Fadil Inceoglu}
\email{fadil@gfz-potsdam.de}

\author[0000-0003-4726-3994]{Fadil Inceoglu}
\affiliation{GFZ German Research Centre for Geosciences, Potsdam, Germany}

\author{Yuri Y. Shprits}
\affiliation{GFZ German Research Centre for Geosciences, Potsdam, Germany}
\affiliation{Institute for Physics and Astronomy, University of Potsdam, Potsdam, Germany}
\affiliation{Department of Earth, Planetary,  and  Space  Science,  University  of  California,  Los  Angeles,  Los  Angeles,  CA, USA}

\author{Stephan G. Heinemann}
\affiliation{Max-Planck-Institute for Solar System Research, Goettingen, Germany}

\author{Stefano Bianco}
\affiliation{GFZ German Research Centre for Geosciences, Potsdam, Germany}




\begin{abstract}

Through its magnetic activity, the Sun governs the conditions in Earth's vicinity, creating space weather events, which have drastic effects on our space- and ground-based technology. One of the most important solar magnetic features creating the space weather is the solar wind, that originates from the coronal holes (CHs). The identification of the CHs on the Sun as one of the source regions of the solar wind is therefore crucial to achieve predictive capabilities. In this study, we used an unsupervised machine learning method, $k$-means, to pixel-wise cluster the passband images of the Sun taken by the Atmospheric Imaging Assembly on {\it the Solar Dynamics Observatory}  (AIA/SDO) in 171 \AA, 193 \AA\,, and 211 \AA\,in different combinations. Our results show that the pixel-wise $k$-means clustering together with systematic pre- and post-processing steps provides compatible results with those from complex methods, such as CNNs. More importantly, our study shows that there is a need for a CH database that a consensus about the CH boundaries are reached by observers independently. This database then can be used as the "ground truth", when using a supervised method or just to evaluate the goodness of the models.

\end{abstract}



\section{Introduction} \label{sec:intro}

The Sun is a magnetically active star that shows various magnetic activity structures extending from its surface to its higher atmospheric layers, such as bipolar active regions (ARs) on the photosphere, filaments in the chromosphere, and coronal holes (CHs) in its corona. Through its magnetic activity, the Sun governs the conditions in the vicinity of Earth and throughout the heliosphere, which creates space weather and space climate. Space weather is defined as the effects of the solar wind, and solar eruptive phenomena, such as flares and coronal mass ejections (CMEs), on Earth's magnetosphere, ionosphere, and thermosphere \citep{2006LRSP....3....2S}. The space weather conditions have drastic effects our space- and ground-based technology \citep{2017RiskA..37..206E}.

One of the most important solar magnetic features creating the space weather and in turn affecting the Earth, is the solar wind. The observations revealed that there are three different types of solar wind; (i) steady fast solar winds originate in the CHs, (ii) unsteady slow winds from opening magnetic loops and active regions, and (iii) transient winds from CMEs \citep{2006LRSP....3....1M}. The identification of the CHs on the Sun as one of the source regions of the solar wind \citep{1968SSRv....8..258W} that creates space weather and in turn influences our space- and ground-based technology is therefore crucial to achieve predictive capabilities.

As the source regions of the steady fast solar winds, CHs are identified as regions of low density collisionless plasma that is generally located above inactive parts of the Sun, where open magnetic field lines extend throughout the heliosphere \citep{2006LRSP....3....2S, 2009LRSP....6....3C}. The magnetic field inside a CH is known to be more unipolar and the CHs show sharp and/or diffuse transition on the boundaries between them and their surroundings \citep{2009LRSP....6....3C}. The temporal evolution of the CH as well as the area they cover on the Sun depends on the solar activity cycle, also known as the Schwabe cycle \citep{1844AN.....21..233S}. During the minimum phase of a solar cycle, the CHs are observed to be larger and located mainly on the solar polar caps. On the inclining phase of a cycle, the CHs are observed to be present at any latitude and to be short-lived. During solar maximum, the CHs are smaller and only exist around mid-latitudes, while on the declining phase of the solar cycle there are more long-lived CHs at lower latitudes and they form closer to the solar equator as the cycle progresses \citep{2020SoPh..295..161H}. Additionally, during the inclining and declining phases of a solar cycle, the CHs can evolve into structures extending from a solar pole to solar equator.

As CHs have lower densities and temperatures, and hence the lowest emission in UV and X-ray in comparison to their surrounding environment consisting of active regions and quiet Sun, they appear as dark regions solar images in wavelengths around 194 \AA\,whether they are on-disk or off-limb CHs \citep{2009LRSP....6....3C}. 

Detection of CHs are done by eye-based on He $\text{I}$ 10830 \AA\,near-infrared absorption line triplet \citep{2002SoPh..211...31H}, histogram-based intensity thresholding on 193 \AA\,and 195 \AA\, passband images of the Sun from the Atmospheric Imaging Assembly \citep[AIA;][]{2012SoPh..275...17L} on {\it the Solar Dynamics Observatory} \citep[SDO;][]{2012SoPh..275....3P} and the Extreme Ultraviolet Imaging Telescope \citep[EIT;][]{1995SoPh..162..291D} on {\it the Solar and Heliospheric Observatory} (SOHO), respectively \citep[CHARM;][]{2009SoPh..256...87K}. Additionally, an automated method for detection and segmentation of CHs based on multi-thermal intensity segmentation using 171 \AA, 193 \AA, and 211 \AA\,passband images of the Sun from the AIA/SDO \citep[CHIMERA;][]{2018JSWSC...8A...2G}, and semi-automated method based on intensity threshold that is modulated by the intensity gradient of a CH have been developed \citep[CATCH;][]{2019SoPh..294..144H}.

There are also methods based on supervised and unsupervised machine learning (ML) methods. \citet{2014A&A...561A..29V} developed a set of segmentation procedures based on spatial possibilistic clustering algorithm (SPoCA) to detect CHs in an unsupervised ML fashion. Identified ARs and CHs by this algorithm are uploaded to the event catalogs in the Heliophysics Event Knowledge (HEK) database \citep{2012SoPh..275...67H}. \citet{2018MNRAS.481.5014I} used convolutional neural networks \citep[CNNs;][]{2014arXiv1404.7828S, 2015Natur.521..436L} based on the U-net architecture \citep{10.1007/978-3-319-24574-4_28} to identify CHs on solar images at 193 \AA\,passband images of the Sun from AIA/SDO. They trained their network using binary maps from Kislovodsk Mountain Astronomical Station. Recently, \citet{2021A&A...652A..13J} utilized a progressively growing architecture based CNNs using data from all 7 channels of AIA/SDO (94 \AA, 131 \AA, 171 \AA, 193 \AA, 211 \AA, 304 \AA\,and 335\AA\,) as well as line-of-sight magnetograms from Helioseismic and Magnetic Imager \citep[HMI;][]{2012SoPh..275..207S} on the SDO. For their network, the authors used binary maps from manually reviewed SPoCA-CH \citep{2018mlts.book..365D}.

In this study, we utilize pixel-wise $k$-means algorithm, which is an unsupervised ML method, to detect CHs based on 171 \AA, 193 \AA, and 211 \AA\, passband images from the AIA/SDO. To achieve this objective, we used data from each channel in different combinations, and compared results from each combination to each other as well as to those from CATCH and the HEK data to calculate their performances. We first describe the data used in this study in Section~\ref{sec:data} and explain the analyses and present our results in Section~\ref{sec:analyses_res}. We discuss the results and conclude in Section~\ref{sec:dis_conc}.

\section{Data} \label{sec:data}

\begin{figure*}
\begin{center}
{\includegraphics[width=5.5in]{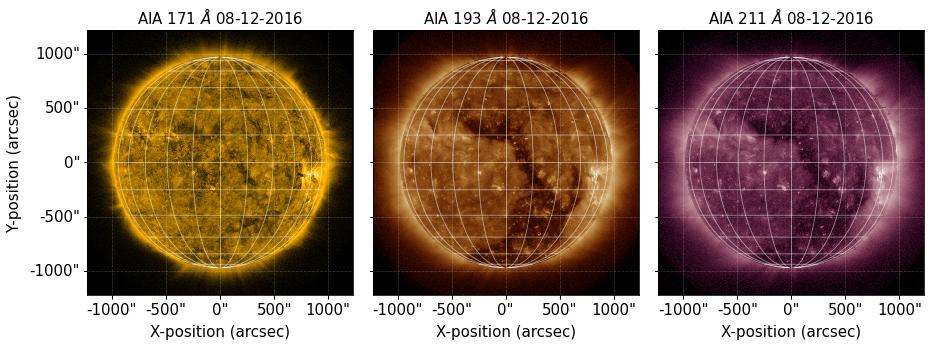}}
\caption{Passband images of the Sun in 171 \AA\,(the left panel), 193 \AA\,(the middle panel), and 211 \AA\,(the right panel) taken by the AIA/SDO on 8 December 2016 at 00:00 UT.}
\label{fig:data}
\end{center}
\end{figure*}

To detect the CHs on the solar corona, we use passband data with 2 second exposure from AIA/SDO in wavelengths 171 \AA, 193 \AA\,, and 211 \AA\,in different combinations (Figure~\ref{fig:data}). The AIA telescope on the SDO takes passband measurements of the Sun in every 12 seconds in full disk with a spatial resolution of 4096$\times$4096 pixels, and each pixel corresponds to 0.6 arsec on the solar disk leading to a spatial resolution of 1.5 arcsec \citep{2012SoPh..275...17L}. These 3 EUV bandpasses are centred on specific spectral emission lines of Fe $\text{IX}$ for 171 \AA, Fe $\text{XII, XXIV}$ for 193 \AA, and Fe $\text{XIV}$ for 211 \AA, which covers the temperature range from $6\times10^{5}$ to $2\times10^{6}$ K, corresponding to the upper transition region, quiet corona (171 \AA), corona and hot flare plasma (193 \AA), and active-region corona (211 \AA)  \citep{2012SoPh..275...17L}.

\section{Analyses and Results} \label{sec:analyses_res}

\subsection{Preprocessing data}

To detect the CHs, we use solar images taken by AIA/SDO in passband images in wavelengths 171 \AA, 193 \AA\,, and 211 \AA\,in different configurations. We also study the most efficient wavelength or configuration of wavelengths to identify the CHs. To achieve this, we compare our CH binary maps with those from the CATCH . We also compared the CH polygons provided by the HEK with the CATCH binary maps to have a base-line with which we compare our results. The CATCH binary maps are selected from the last two months of each year in a time-range from  November 2010 to December 2016, extending through solar cycle 24. The CATCH data in this period is reliable with minimal uncertainties. The total of 237 CATCH CH binary maps consist only contributions from the longitudinal range of $\left[-400, 400\right]$ arcseconds in helioprojective coordinates as in this region the CHs can be identified more robustly \citep{2021A&A...652A..13J}. We also imported CH polygons from the HEK database for the same dates as the CATCH maps, and converted them into binary maps.

\begin{figure*}
\begin{center}
{\includegraphics[width=5.5in]{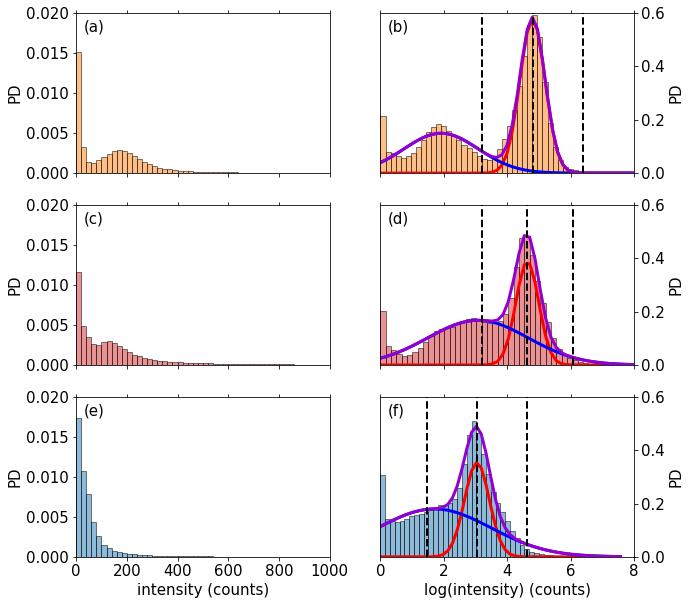}}
\caption{Probability densities of AIA/SDO 171 \AA\,(top panel), 193 \AA\,(middle panel), and 211 \AA\,(bottom panel) intensities of the solar disk on 8 December 2016 at 00:00 UT. The left panels show the probability densities of the preprocessed data, while the right panels show probability densities of the post-processed data. The vertical dashed lines show mean ($\mu$) and $\mu\pm4\sigma$ values calculated to identify the threshold values.}
\label{fig:histograms}
\end{center}
\end{figure*}

In total, we analyze 237 days of data. for each date, we import the level 1 data in 171 \AA, 193 \AA\,, and 211 \AA\,wavelengths and preprocess them using {\it aiapy} \citep{barnes_w_t_2020_4274931,Barnes2020} and {\it SunPy} \citep{sunpy_community2020,stuart_j_mumford_2021_5751998} python packages. This step consists of correcting the data for instrument degradation, for pointing and observer location. Following to these corrections, we registered and aligned the data and normalize it so it has a unit of count/pixel/second. Following these corrections, we correct the passband images for limb brightening using annulus limb brightening correction approach \citep{2014A&A...561A..29V}. We then deconvolve the passband images using instrument point spread function for each wavelength, and rescaled them to 1024$\times$1024 using spline method. As the final step, we log-norm transformed the data.

Following these steps, we created histograms of each data set to determine the lower and upper threshold values. Determining these values allows us to increase the contrast in the data. To avoid using any arbitrary values for these thresholds and to have a more systematic approach for determining these values, we fit a bimodal gaussian curve to each histogram (Figure~\ref{fig:histograms}),  where it is possible. For some dates, however, it was not possible to fit a bimodal gaussian fit. For these dates, we used a unimodal gaussian fit. Using the obtained parameters of the gaussian fits, we calculated the lower- and upper-threshold values based on the mean and standard deviation values of the higher peak (the right panels of Figure~\ref{fig:histograms}), because the lower peak represents the CH pixels \citep{2019SoPh..294..144H}. For each date in the dataset, we calculate a lower-threshold value for each wavelength based on ($\mu - 4\sigma$), while the upper-threshold value is determined based on ($\mu + 4\sigma$). Values below (above) the lower-threshold (upper-threshold) value are stacked to have only one value that is the threshold value.

\begin{figure*}
\begin{center}
{\includegraphics[width=5.5in]{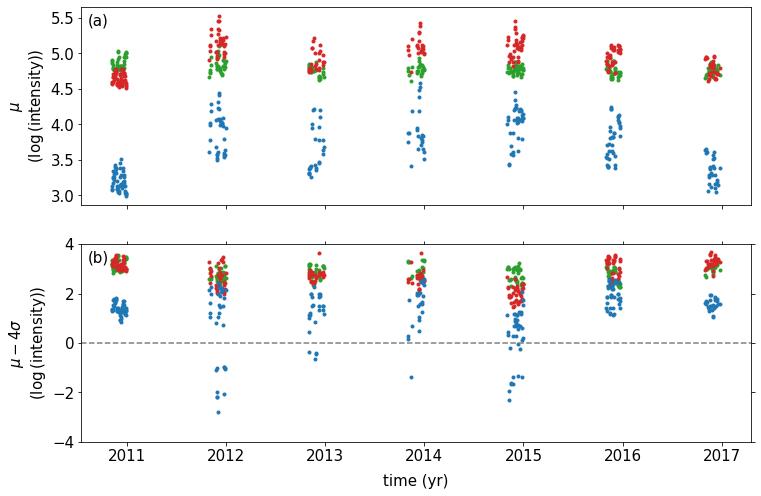}}
\caption{Calculated mean ($\mu$) (a) and lower threshold values ($\mu - 4\sigma$) (b) for AIA/SDO 171 \AA\,(green), 193 \AA\,(red), and 211 \AA\,(blue) passband images for the study period. Note that there are 27 points below zero, meaning that no lower threshold value could be calculated, therefore no thresholding applied to the 211 \AA\,passband data on these dates.}
\label{fig:temporal_threshold}
\end{center}
\end{figure*}

We then investigate the temporal variations in the calculated mean ($\mu$) and the lower threshold values ($\mu - 4\sigma$) (Figure~\ref{fig:temporal_threshold}). The $\mu$ values of 193 \AA\, and 211 \AA\,passband images show variations in phase with the solar cycle, while the $\mu$ values of 171 \AA\,does not show such a trend (Figure~\ref{fig:temporal_threshold}a). The $\mu$ values for each passband images also show day-to-day fluctuations. Similarly, the lower threshold values show day-to-day fluctuations as well. These fluctuations have wider range for the threshold values calculated for the 211 \AA\,passband images especially during the maximum phase of the solar cycle, while the other two channels do not exhibit such wide fluctuations (Figure~\ref{fig:temporal_threshold}b). An important feature to note is the "negative" threshold values found for the 211 \AA\,passband images. There are 27 days where the lower thresholds are negative values. However, as this does not have a physical meaning, the threshold values for these days were accepted as zero. The reason for the negative values come from the underlying shape of the gaussians.

\subsection{Pixel-wise clustering the images using the $k$-means algorithm}

After increasing the contrast in each image based on their individual mean and standard deviation values, we created 4 different data sets; (i) 193 \AA\, image, (ii) 211 \AA\, image, (iii) 193\AA\,and 211\AA\, composite image (2 channel composite, 2CC), and (iv) 171 \AA, 193\AA\,, and 211\AA\, composite image (3 channel composite, 3CC). We then pixel-wise cluster each image using the $k$-means method. This method is used to automatically cluster a given data set into $k$ groups of equal variance \citep{macqueen1967}. The most commonly used clustering criterion is the sum of squared Euclidian distances (SSD), also known as the within-cluster sum-of-squares, of each data point to centroid of the cluster, to which that data point is attained \citep{LIKAS2003451}. The $k$-means algorithm first randomly selects $k$ cluster centroids, and then iteratively refine these initial cluster centroids by assigning each Euclidian distance to its closest cluster centroid. Then the algorithm updates each cluster centroid value to be the mean of its elements by minimizing the SSD \citep{wagstaff2001constrained,LIKAS2003451}.

\begin{figure}
\begin{center}
{\includegraphics[width=3in]{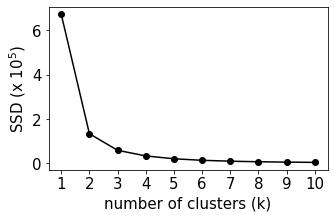}}
\caption{Sum of squared distances (SSD) calculated for each number of clusters, which ranges from 1 to 10 for passband data in 193 \AA\,on 8 December 2017 at 00:00 UT.}
\label{fig:scree}
\end{center}
\end{figure}

The number of clusters, the $k$ value, for this method is an input parameter. To choose the optimum number of clusters, we used the scree-plot method \citep{10.1145/2723372.2737793}. In this method, we use k = 1, 2, 3, …, 10 and calculate the the sum of squared distances (SSD) for each $k$ value. The results show that after the cluster number 3, any further decrease in SSD is very small compared to previous ones, which means that the optimum $k$ value to use, is 3 (Figure~\ref{fig:scree}). This indicates that there are darker regions, brighter regions, and regions that surround them, which can be attained to the CHs, active regions, and the quiet Sun.

The $k$-means method allows us to determine a threshold value for single channel inputs, a threshold line for 2 channel inputs, and a threshold surface for 3 channel inputs in a systematic way that enables us to deter from choosing these thresholds arbitrarily. Additionally, this method, when automated, is flexible enough for day-to-day variations in solar images, providing a dynamical response to them.

We calculate segmentation maps for each date using $k$-means method throughout solar cycle 24. Following that, we convert these maps to binary maps by merging the 2 clusters that identify brighter regions (active regions) and regions that surrounds darker and brighter regions (quiet sun). The reason we did not use $k$ value as 2, is to avoid overestimation of the darker pixels on the passband images of the solar disk. We then remove small dotted-like regions using {\it morphology} module of scikit-image package \citep{scikit-image}. This method requires two inputs; the smallest allowable object size and connectivity, which we use 200 and 10 pixels, respectively. We also used morphological closing using a disk-shaped footprint with a radius of 2 pixels to remove smaller holes in identified CHs. The reason for using a smaller footprint is to try to avoid smoothing out larger bright points in identified CHs, which might be related to the Coronal Bright Points \citep{2006ApJ...642..562K, 2014ApJ...796...73H, 2018ApJ...864..165W}.

In addition to the 4 different binary maps types generated based on the 193 \AA, 211 \AA, 2CC, and 3CC, we generated another type of binary map. We generated them based on the overlap between binary maps of the 193 \AA\,and 211 \AA\,images, which we will refer to as the 2 Channel Overlap (2CO). The 2CO binary maps are created if a pixel is simultaneously identified as a CH pixel in the two binary maps from the 193 \AA\,and 211 \AA\,images. Those pixel, which are not simultaneously identified as a CH are then accepted as non-CH pixels.

\subsection{Pixel-wise evaluation metrics}

To calculate the performances of our binary maps generated by the $k$-means method for each date, we used pixel-wise evaluation metrics. As there will be an imbalance between non-CH and CH pixels in the passband and composite images of the Sun, we use intersection over union (IoU), also known as the Jaccard index \citep{https://doi.org/10.1111/j.1469-8137.1912.tb05611.x}, and true skill statistics (TSS) \citep{hanssen1965relationship} as pixel-wise evaluation metrics. To calculate these metrics, we used binary maps from CATCH. IoU and TSS are calculated based on each confusion matrix for each date using;

\begin{eqnarray} 
\label{eq:TSS}
IoU &=& \frac{TP}{TP + FP + FN}, \\
\nonumber \\ 
TSS &=&  \frac{TP}{TP+FN}-\frac{FP}{FP+TN}
\end{eqnarray}

\noindent where TP, TN, FP, and FN denote pixel-wise calculated number of true positives, true negatives, false positives, and false negatives, respectively.

\begin{figure*}
\begin{center}
{\includegraphics[width=5.5in]{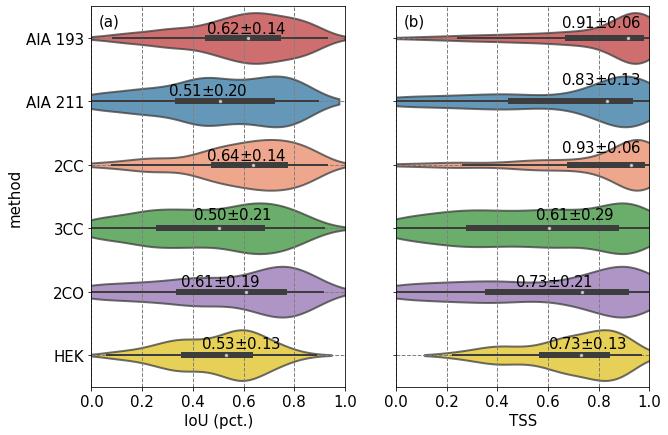}}
\caption{The distributions of the calculated IoU (a) and TSS (b) values between binary maps generated in this study and the CATCH, together with those between the HEK database and the CATCH. The white dots indicate the median value for each distribution. We also show the median values together with median absolute deviation for each evaluation metric in the figure. The red, blue, orange, green, purple, and yellow colors show AIA 193, AIA 211, 2CC, 3CC, 2CO, and HEK binary maps, respectively.}
\label{fig:IoU_TSS}
\end{center}
\end{figure*}

The distributions of the IoU values calculated between our and the CATCH binary maps together with those between the HEK and the CATCH binary maps show that the IoU for the HEK CH binary maps has a median value of 0.53$\pm$0.13, while our results from the AIA 193 and 2CC show median values of 0.62$\pm$0.14 and 0.64$\pm$0.14, respectively. This indicates a better overlap of the identified CHs from our method with those generated by CATCH. The other three binary maps from our study, the AIA 212, 3CC, and 2CO, result in IoU values of 0.51$\pm$0.20, 0.50$\pm$0.21, and 0.61$\pm$0.19, respectively (Figure~\ref{fig:IoU_TSS}a)..

The median TSS values of the AIA 193 and 2CC are 0.91$\pm$0.06 and 0.93$\pm$0.06, respectively (Figure~\ref{fig:IoU_TSS}b), while the median TSS value for the HEK is 0.73$\pm$0.13. These results indicate that our binary maps generated by AIA 193 and 2CC are more in line with those from CATCH. The AIA 212, 3CC, and 2CO, show median TSS values lower than AIA 193 and 2CC (Figure~\ref{fig:IoU_TSS}b).

\subsection{Coronal hole areas}

To further validate our results against the HEK and CATCH results, we calculate the total areas of the CHs on the solar disk in percentage of CH coverage on the solar disk. To achieve this, we first corrected each pixel in our binary maps for projection effects by applying;

\begin{eqnarray} 
\label{eq:TSS}
A_i &=& \frac{A_{i, proj}}{\cos\alpha_i},
\end{eqnarray}

\noindent where A$_i$ and $\alpha_i$ denote the corrected pixel area and the heliographic angular distance of each pixel to the center of the solar disk as seen from the AIA/SDO, respectively.

\begin{figure*}
\begin{center}
{\includegraphics[width=5.5in]{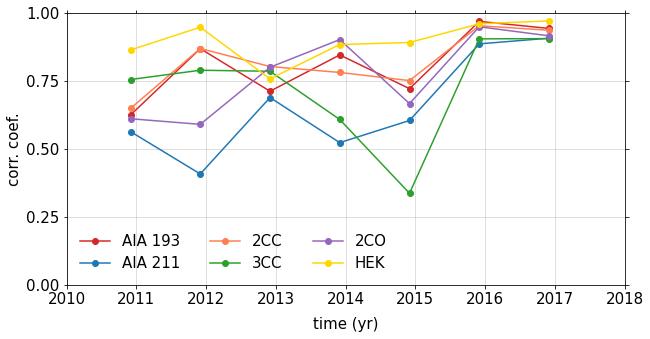}}
\caption{Temporal evolution of the correlation coefficients between total CH areas from our method, HEK against CATCH data through November 2010 and December 2016, extending through solar cycle 24. Note that the correlations are calculated using data during the last two months of each year (see text).}
\label{fig:corr_tseries}
\end{center}
\end{figure*}

We calculated the Pearson correlation coefficients for each year between results from our study, HEK binary maps and CATCH (Figure~\ref{fig:corr_tseries}). We need note that we use the last two months of each year to calculate the correlations. Similar to the results obtained for IoU and TSS, AIA 193 and 2CC generally provide higher correlations through the study period. Interestingly after 2014, the correlation coefficients calculated for every binary map become similar and evolve in parallel until 2016 (Figure~\ref{fig:corr_tseries}).

\begin{figure*}
\begin{center}
{\includegraphics[width=5.5in]{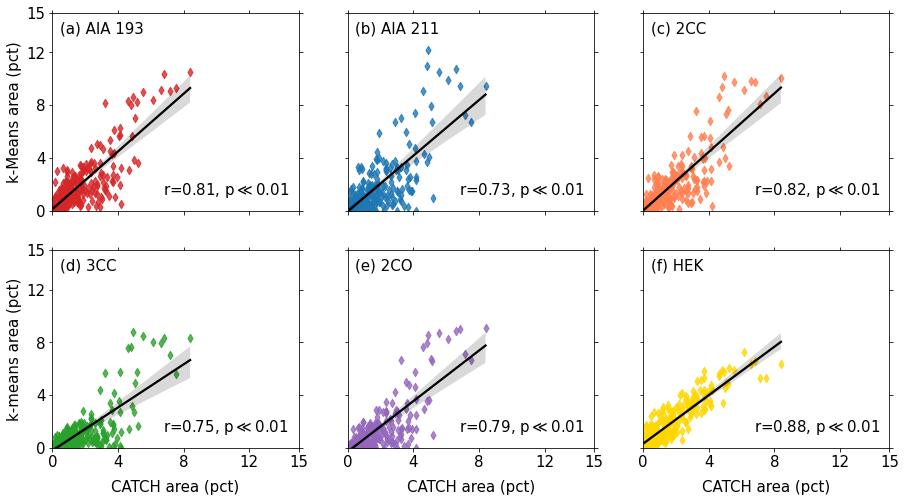}}
\caption{The total percentage areas from this study (a to e) and HEK data base (f) as a function of the areas from CATCH. The black solid lines show the linear fits, while the shaded areas show uncertainty. We also show the Pearson correlation coefficients and their statistical significances. The color coding is the same in Figure~\ref{fig:IoU_TSS}.}
\label{fig:areas}
\end{center}
\end{figure*}

We also calculated the overall correlations between the binary maps from our study and HEK, and binary maps from CATCH. The highest correlation of 0.88 for the CH areas is observed between the HEK and the CATCH data, while our 2CC gave a correlation coefficient of 0.82, followed closely by AIA 193 that gave a correlation coefficient of 0.81. The correlation coefficients for the 2CO, 3CC, and AIA 212 are 0.79, 0.75, and 0.73 respectively (Figure~\ref{fig:areas}).

\subsection{Comparison of the CH binary maps}

\begin{figure*}
\begin{center}
{\includegraphics[width=3.5in]{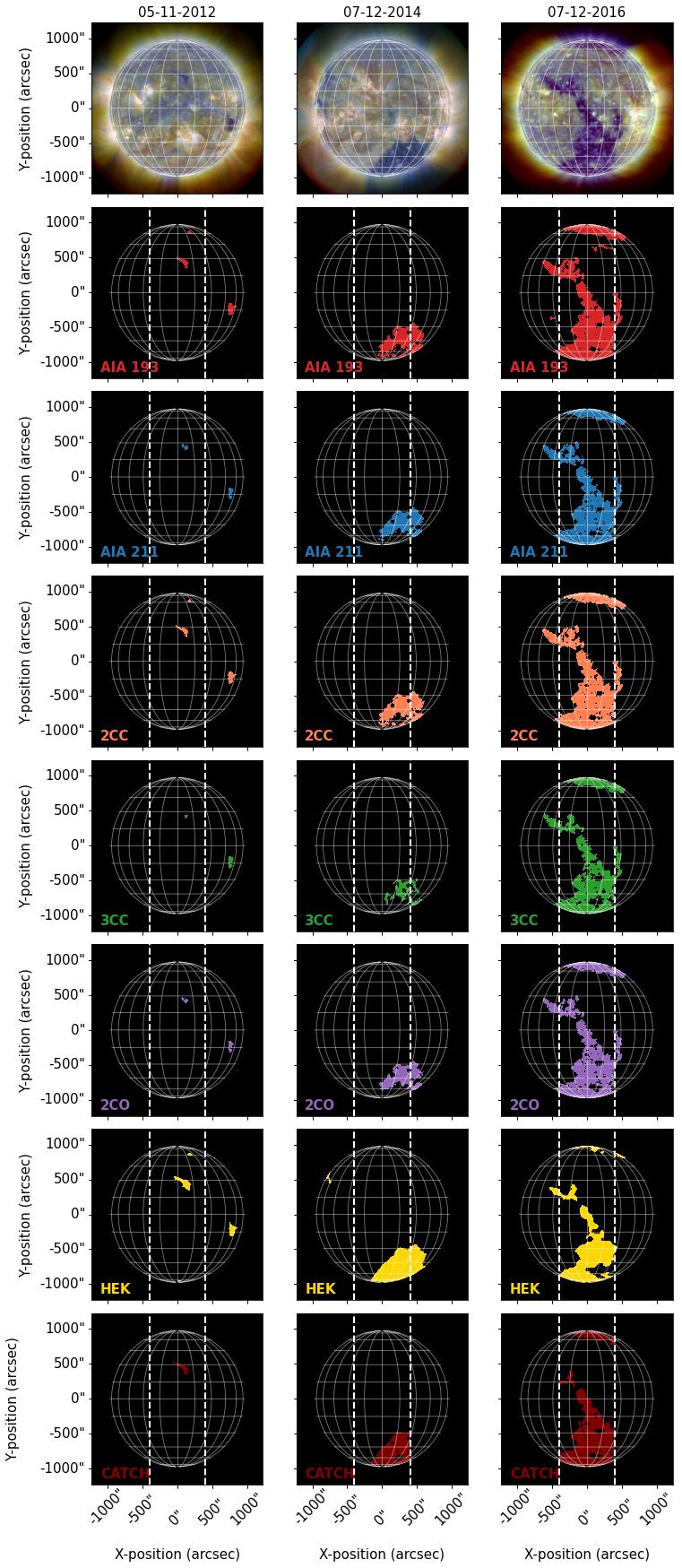}}
\caption{The CH binary maps for 05 November 2012 (top row), 07 December 2014 (middle row), and 07 December 2016 (bottom row) identified from the AIA 193, AIA 211, 2CC, 3CC, 2CO together with binary maps from the HEK and CATCH. The vertical white dashed lines indicate the longitudinal range of $\left[-400, 400\right]$ arcseconds in helioprojective coordinates. The color coding is the same in Figure~\ref{fig:IoU_TSS}.}
\label{fig:three_dates}
\end{center}
\end{figure*}

We then select three dates that represent different phases of solar cycle 24 to compare the CH binary maps. These dates are (i) 05 November 2012 on the inclining phase before the cycle maximum, (ii)  07 December 2014 right after the solar cycle maximum, and (iii) 07 December 2016 on the declining phase of solar cycle 24 (Figure~\ref{fig:three_dates}).

On the inclining phase of solar cycle 24, on 05 November 2012, our method identifies smaller CHs. The results from the AIA 193, 3CC, and 2CO are observed to be more in line with those from the CATCH, where there is only one CH at $\left[0, 500\right]$ arcseconds in helioprojective coordinates. The results from the AIA 211 and the 2CC, on the other hand, more in line with those from the HEK database (the top row of Figure~\ref{fig:three_dates}). On 07 December 2014, a few months after the cycle maximum, the binary maps from the AIA 193, the 3CC, and the 2CO show similar CH coverage on the solar disk to the CATCH within the longitudinal range of $\left[-400, 400\right]$ arcseconds. All of the CH binary maps from our method, except for the 3CC, are similar to the CHs from the HEK showing a small coronal hole near $\left[-750, 500\right]$ arcseconds (the middle row of Figure~\ref{fig:three_dates}). On the declining phase of solar cycle 24, on 07 December 2016, the CH areas identified using the AIA 193, the 2CC, and the 3CC are in line with those from the HEK database and CATCH. On this date, the total CH area coverage also reaches its maximum, where it extends from the southern solar pole to the solar equator (the bottom row of Figure~\ref{fig:three_dates}).

\begin{figure*}
\begin{center}
{\includegraphics[width=6.5in]{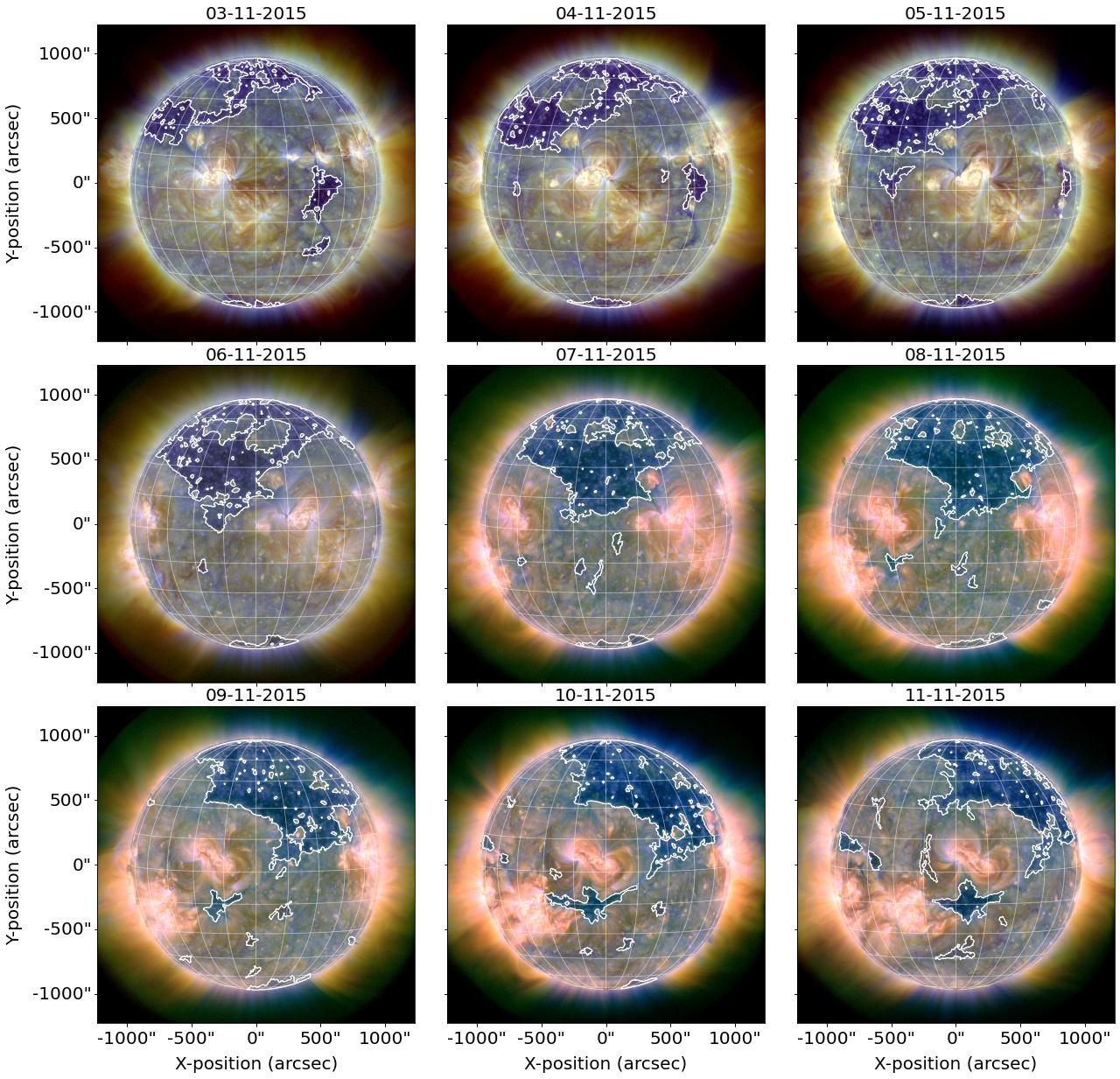}}
\caption{The CH binary maps for a time sequence from 03 through 11 November 2015 identified from the 2CC.}
\label{fig:2cc_seq}
\end{center}
\end{figure*}

To evaluate the consistency of our results, we plotted the detected CHs using 2CC on the dates from 3 November 2015 through 11 November 2015 (Figure~\ref{fig:2cc_seq}). The temporal evolution of the detected CHs close to the solar equator is consistent with the solar rotation. Formation and evolution of a new CH, again close to the solar equator, starting from the 6th of November through 11th of November can also be observed. In addition, temporal evolution of the large CH on the northern solar hemisphere is also consistent in each date (Figure~\ref{fig:2cc_seq}).

\begin{figure*}
\begin{center}
{\includegraphics[width=4in]{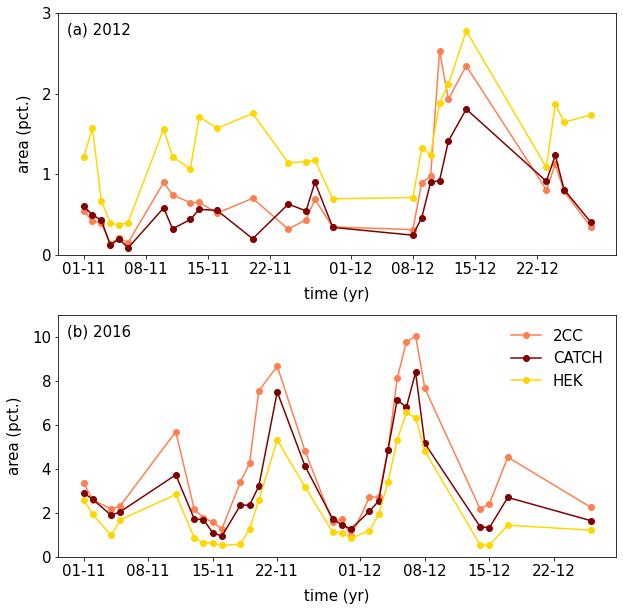}}
\caption{The CH areas during the last two months of 2012 (a) and 2016 (b). The coral, gold, and maroon lines represent 2CC, HEK, and CATCH data, respectively.}
\label{fig:area_temporal}
\end{center}
\end{figure*}

To further investigate the consistency, we checked the day-to-day temporal evolution of the areas during 2012 and 2016 (Figure~\ref{fig:area_temporal}). Note that the areas are calculated for the last two months of each year. In 2012, there is a general good agreement between our 2CC, CATCH, and HEK CHs especially during December, whereas in November, the HEK CH areas are larger compared to our 2CC and the CATCH (Figure~\ref{fig:area_temporal}a). During 2016, on the other hand, CH areas from the three sources covary with some small differences in amplitudes (Figure~\ref{fig:area_temporal}b).

\section{Discussion and Conclusions} \label{sec:dis_conc}

CHs are the source regions of the steady fast solar winds, which results in CIR driven storms, the so-called HILDCAA events \citep{TSURUTANI1987405}. In comparison to their surroundings, CH have lower plasma density and temperatures and therefore they have the lowest emissions in UV and X-ray wavelength range. This physical feature makes them appear as darker regions in passband images of the Sun taken in these wavelengths. CHs are also known to have very complex magnetic structures extending from the photosphere to the corona \citep{2018ApJ...863...29H, 2021SoPh..296..141H}, where the open magnetic field lines extend into the interplanetary medium. They also show solar cycle dependence.

There are several methods to identify CHs on the solar images taken by AIA/SDO and EIT/SOHO based on histograms \citep{2009SoPh..256...87K}, multi-thermal intensity segmentation \citep{2018JSWSC...8A...2G}, and intensity threshold, which is modulated by the intensity gradient of a CH \citep{2019SoPh..294..144H}. Recently, unsupervised and supervised ML methods are used to detect CHs using single or multi-channel passband data from the AIA/SDO \citep{2014A&A...561A..29V,2018MNRAS.481.5014I,2021A&A...652A..13J}. The supervised ML methods mainly rely on the CNNs for image segmentation. These methods, however, require a reliable training data set that is CH polygons detected either by an observer or by an unsupervised method.

In our study, to identify the CHs we used a simple clustering algorithm, $k$-means, to pixel-wise cluster the passband images of the Sun taken in 171 \AA, 193 \AA, and 211 \AA\, by the AIA/SDO covering the time period between November 2010 and December 2016. In addition to using a single-channel approach, we used different combinations of these channels. To detect the lower and upper threshold values, we fitted bimodal gaussians to the probability densities of intensities for each channel on each date. We then calculated the thresholds based on the mean and standard deviation of the local maximum at higher intensities. To cluster the passband images, we used the $k$-means method, where the optimum number of clusters, 3, is calculated based on the scree plot. The $k$-means method, together with pre- and post-processing steps enabled us to build a automated flexible approach which dynamically responds to day-to-day variations in solar images. As a result we obtained 5 different binary maps for each identified CHs, that are  (i) AIA 193, (ii) AIA 211, (iii) 2CC, (iv) 3CC, and (v) 2CO. We then calculated pixel-wise evaluation metrics based on CH binary maps from CATCH and compared our results with each other as well as those from the HEK database. Following that, we calculated the total percentage area identified as a CH per date, after correcting the binary maps for the projection effects. 

Our results show that the 2CC, a composite image using only 193 and 211 \AA\ passband images, provides the best results that is closely followed by results from AIA 193. The median IoU and TSS values for the 2CC are 0.64$\pm$0.14 and 0.93$\pm$0.06, respectively, while they are 0.62$\pm$0.14 and 0.91$\pm$0.06 for the AIA 193. Our results show higher similarity to CATCH results than the HEK database (IoU = 0.53$\pm$0.13 and TSS = 0.73$\pm$0.13). Our results provided better overlap with the CATCH data than those obtained by the CHRONNOS method \citep{2021A&A...652A..13J} for the same period, which provided mean IoU and TSS values as as 0.63 and 0.81, respectively. This method uses all of the 7 channels from the AIA/SDO and line-of-sight magnetograms from the HMI/SDO in a progressively growing CNNs \citep{2021A&A...652A..13J}. Even though our results from AIA 193 and 2CC also provide high overall correlations, they are still lower than the correlation coefficient of 0.88 between the HEK binary maps and CATCH.

We also showed the consistency of our results, especially from the 2CC method ,when the formation and temporal evolution of the CHs are considered. Our method was able to identify and track the CHs from the 3 November through 11 November for 9 consecutive days. Additionally, temporal variations of CH areas from our method follows the trends that is observed in the CATCH and HEK CH areas.

To investigate the effects of the chosen lower and upper threshold values, we also calculated the same evaluation metrics and areas for the threshold ranges of {$\mu\pm3\sigma$}, $\mu\pm5\sigma$, as well as for cases where we do not apply any thresholding at all. Similarly, we calculated the thresholds based on the bimodal gaussian fit and the mean and standard deviation of the local maximum at the higher intensities. However, using different thresholds, and also not using any thresholds, provided lower evaluation metrics as well as correlation coefficients of the total areas.

Interestingly enough, our results show significant discrepancies between the identified CHs using our method, HEK and CATCH when we look at the temporal variations in the correlation coefficients calculated for the total areas. Recently, some steps have been taken to create a reliable database where there is a consensus about the CH boundaries and their uncertainties are being discussed \citep{2021ApJ...918...21L,2021ApJ...913...28R}.

In conclusion, as an unsupervised ML method, using the $k$-means clustering provides better results with those from complex methods, such as CNNs. One of the most important steps in this method is the preprocessing of the data and the choice of the lower and upper threshold values in a more systematic way, which then can lead to automation of the CH detection at any given date or a date range. More importantly, our study shows that there is need for a CH database that a consensus about the CH boundaries are reached by observers independently, and that can be used as the "ground truth", when using a supervised method or just to evaluate the goodness of the models.

\acknowledgments

This research is supported by the Helmholtz Imaging Platform, Solar Image-based Modelling (SIM) ZT-I-PF4-016.



\bibliography{paper_bibliography}{}

\begin{thebibliography}{}
\expandafter\ifx\csname natexlab\endcsname\relax\def\natexlab#1{#1}\fi
\providecommand{\url}[1]{\href{#1}{#1}}
\providecommand{\dodoi}[1]{doi:~\href{http://doi.org/#1}{\nolinkurl{#1}}}
\providecommand{\doeprint}[1]{\href{http://ascl.net/#1}{\nolinkurl{http://ascl.net/#1}}}
\providecommand{\doarXiv}[1]{\href{https://arxiv.org/abs/#1}{\nolinkurl{https://arxiv.org/abs/#1}}}

\bibitem[{Barnes {et~al.}(2020{\natexlab{a}})Barnes, Cheung, Bobra, Boerner,
  Chintzoglou, Leonard, Mumford, Padmanabhan, Shih, Shirman, Stansby, \&
  Wright}]{barnes_w_t_2020_4274931}
Barnes, W., Cheung, M., Bobra, M., {et~al.} 2020{\natexlab{a}}, {aiapy: A
  Python Package for Analyzing Solar EUV Image Data from AIA}, v0.3.1,  Zenodo,
  \dodoi{10.5281/zenodo.4274931}

\bibitem[{Barnes {et~al.}(2020{\natexlab{b}})Barnes, Cheung, Bobra, Boerner,
  Chintzoglou, Leonard, Mumford, Padmanabhan, Shih, Shirman, Stansby, \&
  Wright}]{Barnes2020}
Barnes, W.~T., Cheung, M. C.~M., Bobra, M.~G., {et~al.} 2020{\natexlab{b}},
  Journal of Open Source Software, 5, 2801, \dodoi{10.21105/joss.02801}

\bibitem[{{Cranmer}(2009)}]{2009LRSP....6....3C}
{Cranmer}, S.~R. 2009, Living Reviews in Solar Physics, 6, 3,
  \dodoi{10.12942/lrsp-2009-3}

\bibitem[{{Delaboudini{\`e}re} {et~al.}(1995){Delaboudini{\`e}re}, {Artzner},
  {Brunaud}, {Gabriel}, {Hochedez}, {Millier}, {Song}, {Au}, {Dere}, {Howard},
  {Kreplin}, {Michels}, {Moses}, {Defise}, {Jamar}, {Rochus}, {Chauvineau},
  {Marioge}, {Catura}, {Lemen}, {Shing}, {Stern}, {Gurman}, {Neupert},
  {Maucherat}, {Clette}, {Cugnon}, \& {van Dessel}}]{1995SoPh..162..291D}
{Delaboudini{\`e}re}, J.~P., {Artzner}, G.~E., {Brunaud}, J., {et~al.} 1995,
  \solphys, 162, 291, \dodoi{10.1007/BF00733432}

\bibitem[{{Delouille} {et~al.}(2018){Delouille}, {Hofmeister}, {Reiss},
  {Mampaey}, {Temmer}, \& {Veronig}}]{2018mlts.book..365D}
{Delouille}, V., {Hofmeister}, S.~J., {Reiss}, M.~A., {et~al.} 2018, {''Chapter
  15 - Coronal Holes Detection Using Supervised Classification}, 365--395,
  \dodoi{10.1016/B978-0-12-811788-0.00015-9}

\bibitem[{{Eastwood} {et~al.}(2017){Eastwood}, {Biffis}, {Hapgood}, {Green},
  {Bisi}, {Bentley}, {Wicks}, {McKinnell}, {Gibbs}, \&
  {Burnett}}]{2017RiskA..37..206E}
{Eastwood}, J.~P., {Biffis}, E., {Hapgood}, M.~A., {et~al.} 2017, Risk
  Analysis, 37, 206, \dodoi{10.1111/risa.12765}

\bibitem[{{Garton} {et~al.}(2018){Garton}, {Gallagher}, \&
  {Murray}}]{2018JSWSC...8A...2G}
{Garton}, T.~M., {Gallagher}, P.~T., \& {Murray}, S.~A. 2018, Journal of Space
  Weather and Space Climate, 8, A02, \dodoi{10.1051/swsc/2017039}

\bibitem[{Hanssen \& Kuipers(1965)}]{hanssen1965relationship}
Hanssen, A., \& Kuipers, W. 1965, On the relationship between the frequency of
  rain and various meteorological parameters (with reference to the problem of
  objective forecasting). (Koninklijk Nederlands Meteorologisch Instituut)

\bibitem[{{Harvey} \& {Recely}(2002)}]{2002SoPh..211...31H}
{Harvey}, K.~L., \& {Recely}, F. 2002, \solphys, 211, 31,
  \dodoi{10.1023/A:1022469023581}

\bibitem[{{Heinemann} {et~al.}(2018){Heinemann}, {Hofmeister}, {Veronig}, \&
  {Temmer}}]{2018ApJ...863...29H}
{Heinemann}, S.~G., {Hofmeister}, S.~J., {Veronig}, A.~M., \& {Temmer}, M.
  2018, \apj, 863, 29, \dodoi{10.3847/1538-4357/aad095}

\bibitem[{{Heinemann} {et~al.}(2019){Heinemann}, {Temmer}, {Heinemann},
  {Dissauer}, {Samara}, {Jer{\v{c}}i{\'c}}, {Hofmeister}, \&
  {Veronig}}]{2019SoPh..294..144H}
{Heinemann}, S.~G., {Temmer}, M., {Heinemann}, N., {et~al.} 2019, \solphys,
  294, 144, \dodoi{10.1007/s11207-019-1539-y}

\bibitem[{{Heinemann} {et~al.}(2021){Heinemann}, {Temmer}, {Hofmeister},
  {Stojakovic}, {Gizon}, \& {Yang}}]{2021SoPh..296..141H}
{Heinemann}, S.~G., {Temmer}, M., {Hofmeister}, S.~J., {et~al.} 2021, \solphys,
  296, 141, \dodoi{10.1007/s11207-021-01889-z}

\bibitem[{{Hewins} {et~al.}(2020){Hewins}, {Gibson}, {Webb}, {McFadden},
  {Kuchar}, {Emery}, \& {McIntosh}}]{2020SoPh..295..161H}
{Hewins}, I.~M., {Gibson}, S.~E., {Webb}, D.~F., {et~al.} 2020, \solphys, 295,
  161, \dodoi{10.1007/s11207-020-01731-y}

\bibitem[{{Hong} {et~al.}(2014){Hong}, {Jiang}, {Yang}, {Bi}, {Li}, {Yang}, \&
  {Yang}}]{2014ApJ...796...73H}
{Hong}, J., {Jiang}, Y., {Yang}, J., {et~al.} 2014, \apj, 796, 73,
  \dodoi{10.1088/0004-637X/796/2/73}

\bibitem[{{Hurlburt} {et~al.}(2012){Hurlburt}, {Cheung}, {Schrijver}, {Chang},
  {Freeland}, {Green}, {Heck}, {Jaffey}, {Kobashi}, {Schiff}, {Serafin},
  {Seguin}, {Slater}, {Somani}, \& {Timmons}}]{2012SoPh..275...67H}
{Hurlburt}, N., {Cheung}, M., {Schrijver}, C., {et~al.} 2012, \solphys, 275,
  67, \dodoi{10.1007/s11207-010-9624-2}

\bibitem[{{Illarionov} \& {Tlatov}(2018)}]{2018MNRAS.481.5014I}
{Illarionov}, E.~A., \& {Tlatov}, A.~G. 2018, \mnras, 481, 5014,
  \dodoi{10.1093/mnras/sty2628}

\bibitem[{Jaccard(1912)}]{https://doi.org/10.1111/j.1469-8137.1912.tb05611.x}
Jaccard, P. 1912, New Phytologist, 11, 37,
  \dodoi{https://doi.org/10.1111/j.1469-8137.1912.tb05611.x}

\bibitem[{{Jarolim} {et~al.}(2021){Jarolim}, {Veronig}, {Hofmeister},
  {Heinemann}, {Temmer}, {Podladchikova}, \& {Dissauer}}]{2021A&A...652A..13J}
{Jarolim}, R., {Veronig}, A.~M., {Hofmeister}, S., {et~al.} 2021, \aap, 652,
  A13, \dodoi{10.1051/0004-6361/202140640}

\bibitem[{{Karachik} {et~al.}(2006){Karachik}, {Pevtsov}, \&
  {Sattarov}}]{2006ApJ...642..562K}
{Karachik}, N., {Pevtsov}, A.~A., \& {Sattarov}, I. 2006, \apj, 642, 562,
  \dodoi{10.1086/500820}

\bibitem[{{Krista} \& {Gallagher}(2009)}]{2009SoPh..256...87K}
{Krista}, L.~D., \& {Gallagher}, P.~T. 2009, \solphys, 256, 87,
  \dodoi{10.1007/s11207-009-9357-2}

\bibitem[{{Lecun} {et~al.}(2015){Lecun}, {Bengio}, \&
  {Hinton}}]{2015Natur.521..436L}
{Lecun}, Y., {Bengio}, Y., \& {Hinton}, G. 2015, \nat, 521, 436,
  \dodoi{10.1038/nature14539}

\bibitem[{{Lemen} {et~al.}(2012){Lemen}, {Title}, {Akin}, {Boerner}, {Chou},
  {Drake}, {Duncan}, {Edwards}, {Friedlaender}, {Heyman}, {Hurlburt}, {Katz},
  {Kushner}, {Levay}, {Lindgren}, {Mathur}, {McFeaters}, {Mitchell}, {Rehse},
  {Schrijver}, {Springer}, {Stern}, {Tarbell}, {Wuelser}, {Wolfson}, {Yanari},
  {Bookbinder}, {Cheimets}, {Caldwell}, {Deluca}, {Gates}, {Golub}, {Park},
  {Podgorski}, {Bush}, {Scherrer}, {Gummin}, {Smith}, {Auker}, {Jerram},
  {Pool}, {Soufli}, {Windt}, {Beardsley}, {Clapp}, {Lang}, \&
  {Waltham}}]{2012SoPh..275...17L}
{Lemen}, J.~R., {Title}, A.~M., {Akin}, D.~J., {et~al.} 2012, \solphys, 275,
  17, \dodoi{10.1007/s11207-011-9776-8}

\bibitem[{Likas {et~al.}(2003)Likas, Vlassis, \& {J. Verbeek}}]{LIKAS2003451}
Likas, A., Vlassis, N., \& {J. Verbeek}, J. 2003, Pattern Recognition, 36, 451,
  \dodoi{https://doi.org/10.1016/S0031-3203(02)00060-2}

\bibitem[{{Linker} {et~al.}(2021){Linker}, {Heinemann}, {Temmer}, {Owens},
  {Caplan}, {Arge}, {Asvestari}, {Delouille}, {Downs}, {Hofmeister}, {Jebaraj},
  {Madjarska}, {Pinto}, {Pomoell}, {Samara}, {Scolini}, \&
  {Vr{\v{s}}nak}}]{2021ApJ...918...21L}
{Linker}, J.~A., {Heinemann}, S.~G., {Temmer}, M., {et~al.} 2021, \apj, 918,
  21, \dodoi{10.3847/1538-4357/ac090a}

\bibitem[{MacQueen(1967)}]{macqueen1967}
MacQueen, J. 1967, in Proceedings of the Fifth Berkeley Symposium on
  Mathematical Statistics and Probability, Volume 1: Statistics (Berkeley,
  Calif.: University of California Press), 281--297.
\newblock \url{https://projecteuclid.org/euclid.bsmsp/1200512992}

\bibitem[{{Marsch}(2006)}]{2006LRSP....3....1M}
{Marsch}, E. 2006, Living Reviews in Solar Physics, 3, 1,
  \dodoi{10.12942/lrsp-2006-1}

\bibitem[{Mumford {et~al.}(2021)Mumford, Freij, Christe, Ireland, Mayer,
  Stansby, Shih, Hughitt, Ryan, Liedtke, Pérez-Suárez, I., Hayes,
  Chakraborty, Inglis, Pattnaik, Sipőcz, Sharma, Leonard, Hewett, Hamilton,
  Manhas, Panda, Earnshaw, Barnes, Choudhary, Kumar, Singh, Chanda, Haque,
  Kirk, Konge, Mueller, Srivastava, Jain, Bennett, Baruah, Arbolante, Maloney,
  Charlton, Mishra, Chorley, Himanshu, Chouhan, Modi, Mason, Sharma, Naman9639,
  Zivadinovic, Rozo, Bobra, Manley, Paul, Ivashkiv, Chatterjee, Stern, von
  Forstner, Bazán, Jain, Evans, Ghosh, Malocha, Stańczak, SophieLemos, Verma,
  Visscher, Singh, Airmansmith97, Buddhika, Pathak, Alam, Agrawal, Sharma,
  Rideout, Bates, Park, Mishra, Goel, Sharma, Taylor, Cetusic, Reiter, Jacob,
  Inchaurrandieta, Dacie, Dubey, Parkhi, Sidhu, Surve, Eigenbrot, Meszaros,
  Bray, Zahniy, Guennou, Bose, Ankit, Chicrala, J, D'Avella, Ballew,
  Price-Whelan, Robitaille, Augspurger, Murphy, Lodha, Krishan, Pandey, honey,
  Verma, neerajkulk, Williams, Wiedemann, Kothari, mridulpandey, Habib, Molina,
  Mampaey, Streicher, Nomiya, Gomillion, Letts, Bhope, Hill, Keşkek, Ranjan,
  Pereira, Dang, Bankar, Bahuleyan, B, Stevens, Agrawal, nakul shahdadpuri,
  Ghosh, Hiware, yasintoda, Krishna, Lyes, Mangaonkar, Cheung, platipo,
  Buitrago-Casas, Mendero, Dedhia, Wimbish, Calixto, Babuschkin, Schoentgen,
  Mathur, Kumar, Verstringe, Dover, Tollerud, Gyenge, Arias, Mekala, MacBride,
  Das, Mishra, Stone, resakra, Agarwal, Chaudhari, Kustov, Smith, Srikanth,
  Jain, Mehrotra, Gaba, Kannojia, Yadav, Paul, Wilkinson, Caswell, \&
  Murray}]{stuart_j_mumford_2021_5751998}
Mumford, S.~J., Freij, N., Christe, S., {et~al.} 2021, SunPy, v3.0.3,  Zenodo,
  \dodoi{10.5281/zenodo.5751998}

\bibitem[{Paparrizos \& Gravano(2015)}]{10.1145/2723372.2737793}
Paparrizos, J., \& Gravano, L. 2015, in Proceedings of the 2015 ACM SIGMOD
  International Conference on Management of Data, SIGMOD '15 (New York, NY,
  USA: Association for Computing Machinery), 1855–1870,
  \dodoi{10.1145/2723372.2737793}

\bibitem[{{Pesnell} {et~al.}(2012){Pesnell}, {Thompson}, \&
  {Chamberlin}}]{2012SoPh..275....3P}
{Pesnell}, W.~D., {Thompson}, B.~J., \& {Chamberlin}, P.~C. 2012, \solphys,
  275, 3, \dodoi{10.1007/s11207-011-9841-3}

\bibitem[{{Reiss} {et~al.}(2021){Reiss}, {Muglach}, {M{\"o}stl}, {Arge},
  {Bailey}, {Delouille}, {Garton}, {Hamada}, {Hofmeister}, {Illarionov},
  {Jarolim}, {Kirk}, {Kosovichev}, {Krista}, {Lee}, {Lowder}, {MacNeice},
  {Veronig}, \& {Cospar Iswat Coronal Hole Boundary Working
  Team}}]{2021ApJ...913...28R}
{Reiss}, M.~A., {Muglach}, K., {M{\"o}stl}, C., {et~al.} 2021, \apj, 913, 28,
  \dodoi{10.3847/1538-4357/abf2c8}

\bibitem[{Ronneberger {et~al.}(2015)Ronneberger, Fischer, \&
  Brox}]{10.1007/978-3-319-24574-4_28}
Ronneberger, O., Fischer, P., \& Brox, T. 2015, in Medical Image Computing and
  Computer-Assisted Intervention -- MICCAI 2015, ed. N.~Navab, J.~Hornegger,
  W.~M. Wells, \& A.~F. Frangi (Cham: Springer International Publishing),
  234--241

\bibitem[{{Scherrer} {et~al.}(2012){Scherrer}, {Schou}, {Bush}, {Kosovichev},
  {Bogart}, {Hoeksema}, {Liu}, {Duvall}, {Zhao}, {Title}, {Schrijver},
  {Tarbell}, \& {Tomczyk}}]{2012SoPh..275..207S}
{Scherrer}, P.~H., {Schou}, J., {Bush}, R.~I., {et~al.} 2012, \solphys, 275,
  207, \dodoi{10.1007/s11207-011-9834-2}

\bibitem[{{Schmidhuber}(2014)}]{2014arXiv1404.7828S}
{Schmidhuber}, J. 2014, arXiv e-prints, arXiv:1404.7828.
\newblock \doarXiv{1404.7828}

\bibitem[{{Schwabe}(1844)}]{1844AN.....21..233S}
{Schwabe}, H. 1844, Astronomische Nachrichten, 21, 233,
  \dodoi{10.1002/asna.18440211505}

\bibitem[{{Schwenn}(2006)}]{2006LRSP....3....2S}
{Schwenn}, R. 2006, Living Reviews in Solar Physics, 3, 2,
  \dodoi{10.12942/lrsp-2006-2}

\bibitem[{{The SunPy Community} {et~al.}(2020){The SunPy Community}, Barnes,
  Bobra, Christe, Freij, Hayes, Ireland, Mumford, Perez-Suarez, Ryan, Shih,
  Chanda, Glogowski, Hewett, Hughitt, Hill, Hiware, Inglis, Kirk, Konge, Mason,
  Maloney, Murray, Panda, Park, Pereira, Reardon, Savage, Sipőcz, Stansby,
  Jain, Taylor, Yadav, Rajul, \& Dang}]{sunpy_community2020}
{The SunPy Community}, Barnes, W.~T., Bobra, M.~G., {et~al.} 2020, The
  Astrophysical Journal, 890, 68, \dodoi{10.3847/1538-4357/ab4f7a}

\bibitem[{Tsurutani \& Gonzalez(1987)}]{TSURUTANI1987405}
Tsurutani, B.~T., \& Gonzalez, W.~D. 1987, Planetary and Space Science, 35,
  405, \dodoi{https://doi.org/10.1016/0032-0633(87)90097-3}

\bibitem[{van~der Walt {et~al.}(2014)van~der Walt, {S}ch\"onberger,
  {Nunez-Iglesias}, {B}oulogne, {W}arner, {Y}ager, {G}ouillart, {Y}u, \& the
  scikit-image contributors}]{scikit-image}
van~der Walt, S., {S}ch\"onberger, J.~L., {Nunez-Iglesias}, J., {et~al.} 2014,
  PeerJ, 2, e453, \dodoi{10.7717/peerj.453}

\bibitem[{{Verbeeck} {et~al.}(2014){Verbeeck}, {Delouille}, {Mampaey}, \& {De
  Visscher}}]{2014A&A...561A..29V}
{Verbeeck}, C., {Delouille}, V., {Mampaey}, B., \& {De Visscher}, R. 2014,
  \aap, 561, A29, \dodoi{10.1051/0004-6361/201321243}

\bibitem[{Wagstaff {et~al.}(2001)Wagstaff, Cardie, Rogers, Schr{\"o}dl,
  {et~al.}}]{wagstaff2001constrained}
Wagstaff, K., Cardie, C., Rogers, S., Schr{\"o}dl, S., {et~al.} 2001, in Icml,
  Vol.~1, 577--584

\bibitem[{{Wilcox}(1968)}]{1968SSRv....8..258W}
{Wilcox}, J.~M. 1968, \ssr, 8, 258, \dodoi{10.1007/BF00227565}

\bibitem[{{Wyper} {et~al.}(2018){Wyper}, {DeVore}, {Karpen}, {Antiochos}, \&
  {Yeates}}]{2018ApJ...864..165W}
{Wyper}, P.~F., {DeVore}, C.~R., {Karpen}, J.~T., {Antiochos}, S.~K., \&
  {Yeates}, A.~R. 2018, \apj, 864, 165, \dodoi{10.3847/1538-4357/aad9f7}

\end{thebibliography}
\bibliographystyle{aasjournal}



\end{document}